\documentclass[aps,prb,showpacs,preprint,superscriptaddress,amsfonts,amssymb,amsmath]{revtex4}

\usepackage{amsfonts}
\usepackage{amsmath}
\usepackage{amssymb}
\usepackage{graphicx}

\usepackage{graphicx}
\begin{document}

\title{The definition of the spin current:
The angular spin current and its physical consequences}
\author{Qing-feng Sun$^{1,*}$ and X. C. Xie$^{2,3}$}
\affiliation{$^1$Beijing National Lab for Condensed Matter Physics
and Institute of Physics, Chinese Academy of Sciences, Beijing
100080, China \\
$^2$Department of Physics, Oklahoma State University, Stillwater,
Oklahoma 74078 \\
$^3$International Center for Quantum structures, Chinese Academy
of Sciences, Beijing 100080, China }
\date{\today}

\begin{abstract}
We find that in order to completely describe the spin transport,
apart from spin current (or linear spin current), one has to
introduce the angular spin current. The two spin currents
respectively describe the translational and rotational motion
(precession)of a spin. The definitions of these spin current
densities are given and their physical properties are discussed.
Both spin current densities appear naturally in the spin
continuity equation. Moreover we predict that the angular spin
current can also induce an electric field $\vec{E}$, and in
particular $\vec{E}$ scales as $1/r^2$ at large distance $r$,
whereas the $\vec{E}$ field generated from the linear spin current
goes as $1/r^3$.
\end{abstract}


\maketitle

\section{Introduction}

Recently, a new sub-discipline of condensed matter physics,
spintronics, is emerging rapidly and generating great
interests.\cite{ref1,ref2} The spin current, the most important
physical quantity in spintronics, has been extensively studied.
Many interesting and fundamental phenomena, such as the spin Hall
effect\cite{ref3,ref4,ref5,yao} and the spin
precession\cite{ref6,ref7} in systems with spin-orbit coupling,
have been discovered and are under further study.

As for the charge current, the definition of the local charge
current density $\vec{j}_e({\bf r},t) = Re [\Psi^{\dagger}({\bf
r},t) e\vec{v} \Psi({\bf r},t)]$ and its continuity equation $
 \frac{d}{dt} \rho^e ({\bf r},t) + \nabla \bullet \vec{j}_e ({\bf r},t) =0
$ is well-known in physics. Here $\Psi({\bf r},t)$ is the
electronic wave function, $\vec{v} = \dot{\bf r}$ is the velocity
operator, and $\rho^e({\bf r},t)=e\Psi^{\dagger}\Psi$ is the
charge density. This continuity equation is the consequence of
charge invariance, i.e. when an electron moves from one place to
another, its charge remains the same. However, in the spin
transport, there are still a lot of debates over what is the
correct definition for spin current.\cite{nn1} The problem stems
from that the spin $\vec{s}$ is no invariant quantity in the spin
transport, so that the conventional defining of the spin current
$<\vec{v}\vec{s}>$ is no conservative. Recently, some studies have
begun investigation in this direction\cite{niu,zhang,addref1},
e.g. a semi-classical description of the spin continuity equation
has been proposed\cite{niu,zhang}, as well as studies
introducing a conserved spin current under special circumstances.\cite{nn1}

In this paper, we study the definition of local spin current
density. We find that due to the spin is vector and it has the
translational and rotational motion, one has to use two
quantities, the linear spin current and the angular spin current,
to completely describe the spin transport. Here the linear spin
current describe the translational motion of a spin, and the
angular spin current is for the rotational motion. The
conventional linear spin current has been extensively studied.
However, the physical meaning of the angular spin current is given
for the first time. The definition of two spin current densities
are given and they appear naturally in the quantum spin continuity
equation. Moreover, we predict that the angular spin current can
generate an electric field similar as with the linear spin
current, and thus contains physical consequences.

The paper is organized as follows. In Section II, we first discuss
the flow of a classical vector. The flow of a quantum spin is investigated
in Section III. In Section IV, we study the problem of electric fields
induced by spin currents.
Finally, a brief summary is given in Section V.

\section{the flow of a classical vector}

Before studying the spin current in a quantum system, we first
consider the classical case. Consider a classical particle having
a vector $\vec{m}$ (e.g. the classical magnetic moment, etc.)
with its magnitude $|\vec{m}|$ fixed under
the particle motion. To completely describe this vector flow (see
Fig.1c), besides the local vector density $\vec{M}({\bf r},t)
=\rho({\bf r},t) \vec{m}({\bf r},t)$ one needs two quantities: the
linear velocity $\vec{v}({\bf r},t)$ and the angular velocity
$\vec{\omega}({\bf r},t)$. Here $\rho({\bf r},t)$ is the particle
density, and $\vec{v}$ and $\vec{\omega}$ describe the
translational and rotational motions, respectively (see Fig.1a and
b). In contrast with the flow of a scalar quantity
in which one only needs one
quantity, namely the local velocity $\vec{v}({\bf r}, t)$, to describe
the translational motion, it is essential to use two quantities,
$\vec{v}({\bf r},t)$ and $\vec{\omega}({\bf r},t)$, for the vector
flow. We emphasize that it is impossible to use one vector to
describe both translational and rotational motions
altogether.

Since $|\vec{m}|$ is a constant, the change of the local vector
$\vec{M}({\bf r},t)$ in the volume element $\Delta V =\Delta
x\Delta y\Delta z$ with the time from $t$ to $t + dt$ can be
obtained (see Fig.1c):
\begin{eqnarray}
 d\left( \vec{M}({\bf r},t)\Delta V \right)  & = &
 \sum\limits_{i=x,y,z} \left[ \frac{\Delta V}{\Delta i}
 v_i({\bf r}, t) dt \vec{M}({\bf r}, t)
 - \frac{\Delta V}{\Delta i}
 v_i({\bf r}+\Delta \hat{i}, t) dt \vec{M}({\bf r}+\Delta \hat{i},
 t) \right] \nonumber \\
 & & + \vec{\omega}({\bf r},t)\times \vec{M}({\bf r},t) \Delta V dt.
\end{eqnarray}
The first and the second terms on the right describe the
classical vector flowing in or out the volume element $\Delta V$
and its rotational motion respectively, and both can cause a
change in the local vector density. When $\Delta V$ goes to
zero, we have the vector continuity equation:\cite{nnote1}
\begin{equation}
 \frac{d}{dt} \vec{M}({\bf r},t) =
  - \nabla \bullet\vec{v}({\bf r},t)\vec{M}({\bf r},t)
  + \vec{\omega}({\bf r},t)\times \vec{M}({\bf r},t),
\end{equation}
where $\vec{v}\vec{M}$ is a tensor, and its element
$(\vec{v}\vec{M})_{ij}=v_i M_j$. Notice that this continuity
equation is from the kinematics and the invariance of $|\vec{m}|$,
in particular it is independent of the dynamic laws.\cite{nnote2}
This is completely same with the continuity equation with
a scalar quantity (e.g. charge). Introducing ${\bf j}_s({\bf r},t)
=\vec{v}({\bf r},t)\vec{M}({\bf r},t)$ and $\vec{j}_{\omega}({\bf
r},t)=\vec{\omega}({\bf r},t)\times\vec{M}({\bf r},t)$, then Eq.(2)
reduces to:
\begin{equation}
 \frac{d}{dt} \vec{M}({\bf r},t) =
  - \nabla \bullet {\bf j}_s({\bf r},t)
  + \vec{j}_{\omega}({\bf r},t).
\end{equation}
Here ${\bf j}_s=\vec{v}\vec{M}$ is from the translational motion
of the classical vector $\vec{m}$, and
$\vec{j}_{\omega}=\vec{\omega}\times\vec{M}$ describes its
rotational motion.\cite{nnote1} Since $\vec{v}$ and $\vec{\omega}$
are called as the linear velocity and the angular velocity
respectively, it is natural to name ${\bf j}_s$ and
$\vec{j}_{\omega}$ as the linear and the angular current
densities.

In order to describe a scalar (e.g. charge) flow, one local
current $\vec{j}_e({\bf r},t)$ is sufficient. Why is it required
to introduce two quantities instead of one to describe a vector
flow? The reason is that the scalar quantity only has the
translational motion, but the vector quantity has two kinds of
motion, the translational and the rotational. So one has to use
two quantities, the velocity $\vec{v}$ and the angular velocity
$\vec{\omega}$, to describe the motion of a single vector.
Correspondingly, two quantities ${\bf j}_s = \vec{v}\vec{M}$ and
$\vec{j}_{\omega}=\vec{\omega}\times\vec{M}$ are necessary to
describe the vector flow.

In the steady state case, the scalar (e.g. charge) continuity
equation reduces into $\nabla \bullet \vec{j}_e =0$, so the scalar
current $\vec{j}_e$ is a conserved quantity. But for a vector
flow, the linear vector current ${\bf j}_s$ is not conserved since
$ \nabla \bullet {\bf j}_s  = \vec{j}_{\omega} $. Whether it is
possible to have a conserved vector current through redefinition?
Of course, this redefined vector current should have a clear
physical meaning and is measurable. In our opinion, this is almost
impossible in the 3-dimensional space. The reasons are: (i) One
can not use a 3-dimensional vector to combine both $\vec{v}$ and
$\vec{\omega}$. Therefore one can also not use a 3-dimensional
tensor to combine both ${\bf j}_s = \vec{v}\vec{M}$ and
$\vec{j}_{\omega}=\vec{\omega}\times\vec{M}$. (ii) Consider an
example, as shown in Fig.3a, a one-dimensional classical vector
flowing along the x-axis. When $x<0$, the vector's direction is in
$+x$ axis. At $0<x<L$, the vector rotates in accompany with its
translational motion. When $x>L$, its direction is along $+y$
axis. Since for $x<0$ and $x>L$ the vector has no rotational
motion, the definition of the vector current is unambiguous, and
the non-zero element is $j_{xx}$ for $x<0$, and $j_{xy}$ for
$x>L$. Therefore, the vector current is obviously different for
$x<0$ and $x>L$, and the vector current is non-conservative.

In our opinion, ${\bf j}_s = \vec{v}\vec{M}$ and
$\vec{j}_{\omega}=\vec{\omega}\times\vec{M}$ already have clear
physical meanings. They also completely and sufficiently describe
a vector flow. One may not need need to enforce a conserved
current. In particular, as shown in the example of Fig.3a,
sometimes it is impossible to introduce a conserved current.

\section{the flow of a quantum spin}

Now we study the electronic spin $\vec{s}$ in the quantum case.
Consider an arbitrary wave function $\Psi({\bf r},t)$. The local
spin density $\vec{s}$ at the position ${\bf r}$ and time $t$ is:
$\vec{s}({\bf r},t)= \Psi^{\dagger}({\bf r},t) \hat{\vec{s}}
\Psi({\bf r},t)$, where $\hat{\vec{s}} =
\frac{\hbar}{2}\hat{\vec{\sigma}}$ with $\hat{\vec{\sigma}}$ being the
Pauli matrices. The time-derivative of $\vec{s}({\bf r},t)$ is:
\begin{equation}
 \frac{d}{dt}\vec{s}({\bf r},t) = \frac{\hbar}{2}
  \left\{
    \left[\frac{d}{dt} \Psi^{\dagger} \right]
    \hat{\vec{\sigma}} \Psi
   + \Psi^{\dagger} \hat{\vec{\sigma}}
    \frac{d}{dt} \Psi \right\}.
\end{equation}
From the Schrodinger equation, we have $\frac{d}{dt}\Psi({\bf
r},t)=\frac{1}{i\hbar}H \Psi({\bf r},t)$ and
$\frac{d}{dt}\Psi^{\dagger}({\bf r},t)=\frac{1}{-i\hbar}[H
\Psi({\bf r},t)]^{\dagger}$. Notice here the transposition in the
symbol $\dagger$ only acts on the spin indexes. By using the above
two equations, the Eq.(4) changes into:
\begin{equation}
  (d/dt) \vec{s}({\bf r},t) =
  [\Psi^{\dagger}\hat{\vec{\sigma}} H\Psi
   - (H\Psi)^{\dagger} \hat{\vec{\sigma}} \Psi ]/2i.
\end{equation}
In the derivation below, we use the following Hamiltonian:
\begin{equation}
  H= \frac{\vec{p}^2}{2m} +V({\bf r}) + \hat{\vec{\sigma}}\bullet
  \vec{B} +\frac{\alpha}{\hbar} \hat{z}\bullet
  (\hat{\vec{\sigma}}\times \vec{p}).
\end{equation}
Note that our results are independent of this specific choice of
the Hamiltonian.\cite{note1} In Eq.(6) the 1st and 2nd terms are
the kinetic energy and potential energy. The 3rd term is the
Zeeman energy due to a magnetic field, and the last term is the
Rashba spin-orbit coupling\cite{ref8,ref9}, which has been
extensively studied recently\cite{ref4,ref6,ref7}. Next we
substitute the Hamiltonian Eq.(6) into Eq.(5), and the Eq.(5)
reduces to:
\begin{eqnarray}
  \frac{d}{dt}\vec{s} & = & -\frac{\hbar}{2}\nabla\bullet
   Re\{\Psi^{\dagger} [\frac{\vec{p}}{m}
   +\frac{\alpha}{\hbar}(\hat{z}\times\hat{\vec{\sigma}})]
   \hat{\vec{\sigma}}\Psi \} \nonumber\\
 & +& Re\{ \Psi^{\dagger} [\vec{B}
  +\frac{\alpha}{\hbar}\vec{p}\times\hat{z}]\times\hat{\vec{\sigma}}
  \Psi \}.
\end{eqnarray}
Introducing a tensor ${\bf j}_s({\bf r},t)$ and a vector
$\vec{j}_{\omega}({\bf r},t)$:
\begin{eqnarray}
  {\bf j}_s({\bf r},t) &=& Re\{\Psi^{\dagger}
   [\frac{\vec{p}}{m}+\frac{\alpha}{\hbar}(\hat{z}\times\hat{\vec{\sigma}})]
    \hat{\vec{s}}\Psi \}\\
  \vec{j}_{\omega}({\bf r},t) &=& Re \{\Psi^{\dagger}
    \frac{2}{\hbar} [\vec{B} +
    \frac{\alpha}{\hbar}(\vec{p}\times\hat{z})]\times\hat{\vec{s}}\Psi
    \},
\end{eqnarray}
then Eq.(7) reduces to:
\begin{eqnarray}
  \frac{d}{dt} \vec{s}({\bf r},t) =
   -\nabla \bullet {\bf j}_s({\bf  r}, t)
    + \vec{j}_{\omega}({\bf r}, t),
\end{eqnarray}
or it can also be rewritten in the integral form:
\begin{equation}
  \frac{d}{dt} \iiint_V \vec{s} dV =
   - \oint_S d\vec{S} \bullet {\bf j}_s
    + \iiint_V \vec{j}_{\omega} dV.
\end{equation}
Due to the fact that $\hat{\vec{v}}=\frac{d}{dt}{\bf r}
=\frac{\vec{p}}{m} +\frac{\alpha}{\hbar}(\hat{z}\times
\hat{\vec{\sigma}})$ and $\frac{d}{dt}\hat{\vec{\sigma}}
=\frac{1}{i\hbar}[\hat{\vec{\sigma}},H] =\frac{2}{\hbar}[\vec{B} +
\frac{\alpha}{\hbar} \vec{p}\times\hat{z}]\times
\hat{\vec{\sigma}}$, Eqs.(8) and (9) become:
\begin{eqnarray}
  {\bf j}_s({\bf r},t) &=& Re\{\Psi^{\dagger}({\bf r},t)
    \hat{\vec{v}} \hat{\vec{s}}\Psi({\bf r},t) \} \\
  \vec{j}_{\omega}({\bf r},t) &=& Re\{\Psi^{\dagger}
    (d\hat{\vec{s}}/dt) \Psi \} = Re\{ \Psi^{\dagger}
    \hat{\vec{\omega}} \times \hat{\vec{s}} \Psi \},
\end{eqnarray}
where $\hat{\vec{\omega}}\equiv \frac{2}{\hbar} [\vec{B} +
\frac{\alpha}{\hbar}(\vec{p}\times\hat{z})]$ is the angular
velocity operator.\cite{note1}

Eq.(10) is the spin continuity equation, which is same with the
classic vector continuity equation (3) although the derivation
process is very different. In some previous works, this equation
has also been obtained in the semiclassical case.\cite{niu,zhang}
Here we emphasize that this spin continuity equation (10) is the
consequence of invariance of the spin magnitude $|\vec{s}|$, i.e.
when an electron makes a motion, either translation or rotation,
its spin magnitude $|\vec{s}|=\frac{\hbar}{2}$ remains a constant.
And the Eq.(10) should be independent with the force (i.e. the
potential) and the torque, as well the the dynamic law. The two
quantities ${\bf j}_s({\bf r},t)$ and $\vec{j}_{\omega}({\bf
r},t)$ in Eq.(10), which are defined in Eqs.(12,13) respectively,
describe the translational and rotational motion (precession) of a
spin at the location ${\bf r}$ and the time $t$. They will be
named the linear and the angular spin current densities
accordingly\cite{add1note1}, similar as $\vec{v}$ and
$\vec{\omega}$ are called the linear and the angular velocities.
In fact, the linear spin current ${\bf j}_s({\bf r},t)$ is
identical with the conventional spin current investigated in
recent studies\cite{note2}. However, we give the physical meaning
of $\vec{j}_{\omega}({\bf r},t)$ for the first time.

Next, we discuss certain properties of ${\bf j}_s({\bf r}, t)$ and
$\vec{j}_{\omega}({\bf r},t)$. Notice that $\vec{j}_{\omega}({\bf
r},t)$ which describe the rotational motion (precession) of the
spin plays a parallel role in comparison with the conventional
linear spin current ${\bf j}_s({\bf r}, t)$ for the spin
transport. (1) In similar with the classical case, it is necessary
to introduce the two quantities ${\bf j}_s({\bf r},t)$ and
$\vec{j}_{\omega}({\bf r},t)$ to completely describe the motion of
a quantum spin. (2) The linear spin current is a tensor. Its
element, e.g. $j_{s,xy}$, represents an electron moving along the
$x$ direction with its spin in the $y$ direction (see Fig.2a). The
angular spin current $\vec{j}_{\omega}$ is a vector. In Fig.2b,
its element $j_{\omega,x}$ describes the rotational motion of the
spin in the $y$ direction and the angular velocity ${\vec
{\omega}}$ in the $-z$ direction. (3) From the linear spin current
density ${\bf j}_s({\bf r}, t)$, one can calculate (or say how
much) the linear spin current $\vec{I}_s$ flowing through a
surface $S$ (see Fig.2d): $ \vec{I}_s^S =\iint_S d\vec{S}\bullet
{\bf j}_s$. However, the behavior for the angular spin current is
different. From the density $\vec{j}_{\omega}({\bf r}, t)$, it is
meaningless to determine how much the angular spin current flowing
through a surface $S$, because the angular spin current describes
the rotational motion not the movement. On the other hand, one can
calculate the total angular spin current $\vec{I}_{\omega}^V$ in a
volume $V$ from $\vec{j}_{\omega}$: $\vec{I}_{\omega}^V = \iiint_V
\vec{j}_{\omega}({\bf r},t) dV$. (4) It is easy to prove that the
spin currents in the present definitions of Eqs.(12,13) are
invariant under a space coordinate transformation as well the
gauge transformation. (5) If the system is in a steady state,
${\bf j}_s$ and $\vec{j}_{\omega}$ are independent of the time
$t$, and $\frac{d}{dt} \vec{s}({\bf r},t)=0$. Then the spin
continuity equation (10) reduces to: $ \nabla \bullet {\bf j}_s =
\vec{j}_{\omega} $ or $  \oint_S d\vec{S} \bullet {\bf j}_s =
\iiint_V \vec{j}_{\omega} dV $. This means that the total linear
spin current flowing out of a closed surface is equal to the total
angular spin current enclosed. If to further consider a quasi one
dimensional (1D) system (see Fig.2d), then one has $\vec{I}_s^{S'}
- \vec{I}_s^{S} =\vec{I}_{\omega}^{V}$. (6) The linear spin
current density ${\bf j}_s=Re\{\Psi^{\dagger} \vec{v}
\hat{\vec{s}} \Psi\}$ gives both the spin direction and the
direction of spin movement, so it completely describes the
translational motion. However, the angular spin current density,
$\vec{j}_{\omega}=Re\{\Psi^{\dagger}\frac{d\hat{\vec{s}}}{dt}\Psi
\} =Re\{\Psi^{\dagger} \vec{\omega}\times\hat{\vec{s}} \Psi \}$
involves the vector product of $\vec{\omega}\times\hat{\vec{s}}$,
not the tensor $\vec{\omega}\hat{\vec{s}}$. Is it correct or
sufficient to describe the rotational motion? For example, the
rotational motion of Fig.2b with the spin $\vec{s}$ in the $y$
direction and the angular velocity $\vec{\omega}$ in the $-z$
direction is different from the one in Fig.2c in which $\vec{s}$
is in the $z$ direction and $\vec{\omega}$ is in the $y$
direction, but their angular spin currents are completely the
same. Shall we distinguish them? It turns out that the physical
results produced by the above two rotational motions (Fig.2b and
2c) are indeed the same. For instance, the induced electric field
by them is identical since a spin $\vec{s}$ has only the direction
but no size (see detail discussion below). Thus, the vector
$\vec{j}_{\omega}$ is sufficient to describe the rotational
motion, and no tensor is necessary.

Now we give an example of applying the above formulas,
Eqs.(12,13), to calculate the spin currents. Let us consider a
quasi 1D quantum wire having the Rashba spin orbit coupling, and
its Hamiltonian is:
\begin{equation}
 H =  \frac{\vec{p}^2 }{2 m} + V(y,z) + \frac{\sigma_z}{2\hbar}[\alpha(x)  p_x
      +p_x \alpha(x)] + \frac{\hbar^2 k_R^2}{2m} ,
\end{equation}
where $k_R(x) \equiv \alpha(x) m/\hbar^2$. $\alpha(x)=0$ for $x<0$
and $x>L$, and $\alpha(x)\not=0$ while $0<x<L$. The other Rashba
term $-\frac{\alpha}{\hbar} \sigma_x p_z$ is neglected because
z-direction is quantized\cite{ref6}. Let $\Psi$ be a stationary
wave function
\begin{eqnarray}
 \Psi({\bf r}) = \frac{\sqrt{2}}{2} e^{ikx}
   \left( \begin{array}{l}
         e^{-i\int_0^x k_R(x) dx } \\
         e^{i\int_0^x k_R(x) dx }
    \end{array} \right) \varphi(y,z),
\end{eqnarray}
where $\varphi(y,z)$ is the bound state wave function in the
confined $y$ and $z$ directions. $\Psi({\bf r})$ represents the
spin motion as shown in Fig.3a, in which the spin moves along the
$x$ axis, as well the spin precession in the $x$-$y$ plane in the
region $0<x<L$.\cite{ref6,ref7} Using Eqs.(12,13), the spin
current densities of the wave function $\Psi({\bf r})$ are easily
obtained. There are only two non-zero elements of ${\bf j}_s({\bf
r})$:
\begin{eqnarray}
  j_{sxx}({\bf r}) & = & \frac{\hbar^2 k}{2m}
  |\varphi(y,z)|^2  \cos 2 \phi(x) ,\\
  j_{sxy}({\bf r}) &= & \frac{\hbar^2 k}{2m}
  |\varphi(y,z)|^2  \sin 2\phi(x) .
\end{eqnarray}
The non-zero elements of $\vec{j}_{\omega}({\bf r})$ are:
\begin{eqnarray}
  j_{\omega x}({\bf r}) & = & - \frac{ \hbar^2 k k_R(x)}{ m}
  |\varphi(y,z)|^2  \sin 2 \phi(x), \\
  j_{\omega y}({\bf r}) &= & \frac{\hbar^2 k k_R }{m}
  |\varphi(y,z)|^2 \cos 2 \phi(x),
\end{eqnarray}
where $\phi(x) =\int_0^x k_R(x)dx$. Those spin current densities
confirm with the intuitive picture of an electron motion,
precession in the $x$-$y$ plane in $0<x<L$ and movement in the $x$
direction (see Fig.3a).

In particular, in the region of $x<0$ and $x>L$,
$\alpha(x)=k_R(x)=0$, and $\hat{\sigma}$ is a good quantum
number, hence, $\vec{j}_{\omega}=0$. In this case, the
definition of the spin current ${\bf j}_s$ is unambiguous.
However, the spin currents are different in $x<0$ and
$x>L$ except for $\phi(L)=n\pi$ ($n=0,\pm1,\pm2$ ...). This is
clearly seen from Fig.3a. Therefore, through this
example, one can conclude that it is sometimes impossible to define a
conserved spin current. The example of Fig.3a
indeed exists and has been studied
before\cite{ref6,ref7}.

Above discussion shows that the linear spin current ${\bf j}_s = \vec{v}\vec{s}$ and
the angular spin current
$\vec{j}_{\omega}=\vec{\omega}\times\vec{s}$ have clear
physical meanings, representing the translational motion and
the rotational motion (precession) respectively. They
completely describe the flow of a quantum spin. Any physical
effects of the spin currents, such as the induced electric field,
can be expressed by ${\bf j}_s$ and
$\vec{j}_{\omega}$.\cite{nnote3}

\section{spin currents induced electric fields}

Recently, theoretic studies have suggested that the (linear) spin
current can induce an electric field
$\vec{E}$\cite{ref10,addref2,ref11}. Can the angular spin current
also induce an electric field? If so this gives a way of detecting
the angular spin current. Following, we study this question by
using the method of equivalent magnetic charge\cite{griffiths}.
Let us consider a steady-state angular spin current element
$\vec{j}_{\omega} dV$ at the origin. Associated with the spin
$\vec{s}$, there is a magnetic moment (MM) $\vec{m}=g\mu_B
\vec{\sigma} = \frac{2g\mu_B}{\hbar} \vec{s}$ where $\mu_B$ is the
Bohr magneton. Thus, corresponding to $\vec{j}_{\omega}$, there is
also a angular MM current $\vec{j}_{m\omega} dV =
\frac{2g\mu_B}{\hbar} \vec{j}_{\omega} dV$. From above
discussions, we already know that $\vec{j}_{m\omega}$ (or
$\vec{j}_{\omega}$) comes from the rotational motion of a MM
$\vec{m}$ (or $\vec{s}$) (see Fig.2b and 2c), and
$\vec{j}_{m\omega} = \vec{\omega}\times \vec{m}$ (or
$\vec{j}_{\omega} = \vec{\omega}\times \vec{s}$). Under the method
of equivalent magnetic charge, the MM $\vec{m}$ is equivalent to
two magnetic charges: one with magnetic charge $+q$ located at
$\delta \hat{n}_m$ and the other with $-q$ at $-\delta \hat{n}_m$
(see Fig.2e). $\hat{n}_m$ is the unit vector of $\vec{m}$ and
$\delta$ is a tiny length. The angular MM current
$\vec{j}_{m\omega}$ is equivalent to two magnetic charge currents:
one is $\vec{j}_{+q}=\hat{n}_j q\delta |\vec{\omega}| \sin \theta
$ at the location $\delta \hat{n}_m$, the other is
$\vec{j}_{-q}=\hat{n}_j q\delta |\vec{\omega}| \sin \theta$ at
$-\delta \hat{n}_m$ (see Fig.2e), with $\hat{n}_j$ being the unit
vector of $\vec{j}_{m\omega}$ and $\theta$ the angle between
$\vec{\omega}$ and $\vec{m}$. In our previous work,\cite{ref11} we
have given the formulae of the electric field induced by a
magnetic charge current. The electric field induced by
$\vec{j}_{m\omega}dV$ can be calculated by adding the
contributions from the two magnetic charge currents. Let $\delta
\rightarrow 0$, and note that  $2q\delta \rightarrow |\vec{m}|$
and $|\vec{\omega}| |\vec{m}| \sin\theta =|\vec{j}_{m\omega}|$, we
obtain the electric field $\vec{E}_{\omega}$ generated by an
element of the angular spin current $\vec{j}_{\omega}dV$:
\begin{equation}
 \vec{E}_{\omega} = \frac{-\mu_0}{4\pi}\int \frac{  \vec{j}_{m\omega} dV \times
 {\bf r} }{r^3}
     = \frac{-\mu_0 g\mu_B}{h} \int \vec{j}_{\omega} dV \times
  \frac{ {\bf r} }{r^3}
\end{equation}
We also rewrite the electric field $\vec{E}_s$ generated by an
element of the linear spin current using the tensor ${\bf j}_s
$:\cite{ref11}
\begin{equation}
 \vec{E}_s
     = \frac{-\mu_0 g\mu_B}{h} \nabla \times \int {\bf j}_s dV
     \bullet
  \frac{ {\bf r} }{r^3},
\end{equation}
Below we emphasize three points: (i) In the large $r$ case, the
electric field $\vec{E}_{\omega}$ decays as $1/r^2$. Note that the
field from a linear spin current $\vec{E}_s$ goes as $1/r^3$. In
fact, in terms of generating an electric field, the angular spin
current is as effective as a magnetic charge current. (ii) In the
steady-state case, the total electric field $\vec{E}_T=
\vec{E}_{\omega}+\vec{E}_s$ contains the property: $\oint_C
\vec{E}_T \bullet d\vec{l} =0$, where $C$ is an arbitrary close
contour not passing through the region of spin current. However,
for each $\vec{E}_{\omega}$ or $\vec{E}_s$, $\oint_C
\vec{E}_{\omega}\bullet d\vec{l}$ or $\oint_C \vec{E}_s\bullet
d\vec{l}$ can be non-zero. (iii) As mentioned above, a angular
spin current $\vec{j}_{\omega}$ may consist of different
$\vec{\omega}$ and $\vec{s}$ (see Fig.2b and 2c). However, the
resulting electric field only depends on
$\vec{j}_{\omega}=\vec{\omega} \times \vec{s}$. This is because a
spin vector contains only a direction and a magnitude, but not a
spatial size (i.e. the distance $\delta$ approaches to zero). In
the limit $\delta \rightarrow 0$, both magnetic charge currents
$\vec{j}_{\pm q}$ reduce to $\vec{\omega} \times \vec{m}/2$ at the
origin. Therefore, the overall effect of the rotational motion is
only related to $\vec{\omega} \times \vec{m}$, not separately on
$\vec{\omega}$ and $\vec{m}$. Hence it is enough to describe the
spin rotational motion by using a vector $\vec{\omega} \times
\vec{s}$, instead of a tensor $\vec{\omega} \vec{s}$.

Due to the fact that the direction of $\vec{s}$ can change during
the particle motion, the linear spin current density ${\bf j}_s$
is not a conserved quantity. It is always interesting to uncover a
conserved physical quantity from both theoretical and experimental
points of view. Let us apply $\nabla \bullet$ acting on two sides
of Eq.(10), we have:
\begin{eqnarray}
  \frac{d}{dt} \left(\nabla \bullet \vec{s}\right) +
   \nabla \bullet \left(\nabla'\bullet {\bf j}_s
    - \vec{j}_{\omega}\right) =0 ,
\end{eqnarray}
where $\nabla'\bullet {\bf j}_s$ means that $\nabla'$ acts on the
second index of ${\bf j}_s$, i.e. $(\nabla'\bullet {\bf j}_s)_i =
\sum_j \frac{d}{dj} j_{s,ij}$ with $i,j \in (x,y,z)$. Define
$\vec{j}_{\nabla\bullet\vec{s}} = \nabla'\bullet {\bf j}_s -
\vec{j}_{\omega}$, the above equation reduces into:\cite{nnote4}
\begin{eqnarray}
  \frac{d}{dt} \left(\nabla \bullet \vec{s}\right) +
   \nabla \bullet \vec{j}_{\nabla\bullet\vec{s}} = 0 .
\end{eqnarray}
This means that the current $\vec{j}_{\nabla\bullet\vec{s}}$ of
the spin divergence is a conserved quantity in the steady
state case. In fact, $-\nabla \bullet \vec{s}({\bf r}, t)$ represents
an equivalent magnetic charge, so
$\vec{j}_{\nabla\bullet\vec{s}}$ can also be named the magnetic
charge current density. Moreover, the total electric field
produced by ${\bf j}_s$ and $\vec{j}_{\omega}$ can be rewritten
as:
\begin{eqnarray}
 \vec{E}_{T} & = & \vec{E}_s +\vec{E}_{\omega} =
  \frac{\mu_0 g\mu_B}{h} \int \left(\nabla'\bullet{\bf j}_s -\vec{j}_{\omega}\right) dV \times
  \frac{ {\bf r} }{r^3} \nonumber\\
  & = &
  \frac{\mu_0 g\mu_B}{h} \int \vec{j}_{\nabla\bullet\vec{s}} dV \times
  \frac{ {\bf r} }{r^3}.
\end{eqnarray}
So the total electric field $\vec{E}_T$ only depends on the
current $\vec{j}_{\nabla\bullet\vec{s}}$ of the spin divergence.
Note that $\vec{E}_T$ can be measured experimentally in principle.
Through the measurement of $\vec{E}_T({\bf r})$,
$\vec{j}_{\nabla\bullet\vec{s}}$ can be uniquely obtained.

In the following, let us calculate the induced electric fields at
the location ${\bf r}=(x,y,z)$ by the spin currents in the
example of Fig.3a. Substituting the spin currents of
Eqs.(16-19) into Eqs.(20,21) and assuming the transverse sizes of
the 1D wire are much smaller than $\sqrt{y^2+z^2}$, the induced
fields $\vec{E}_{\omega}$ and $\vec{E}_s$ can be obtained
straightforwardly. Then the total electric field $\vec{E}_T
=\vec{E}_{\omega}+\vec{E}_s $ is:
\begin{eqnarray}
  \vec{E}_T & = & a \frac{\hbar k}{m} \nabla \int \frac{z \sin 2\phi(x')}
            {[(x-x')^2+y^2+z^2]^{3/2}} dx' \nonumber\\
            & = &
           a \nabla \int (\vec{\cal{V}}\times\vec{\cal{S}})\bullet
            \frac{{\bf r}-{\bf r'}}{|{\bf r}-{\bf r'}|^3} dx'
\end{eqnarray}
where $\vec{\cal{V}} = (\hbar k/m , 0,0)$, $\vec{\cal{S}} = (\cos
2\phi(x') ,\sin 2\phi(x'), 0)$, ${\bf r'} =(x',0,0)$, the constant
$a= \mu_0 g\mu_B  \rho_s/4\pi $, and $\rho_s$ is the linear
density of moving electrons under the bias of an external voltage.
The total electric field $\vec{E}_T$ represents the one generated
by a 1D wire of electric dipole moment $\vec{p}_e = (0,0, c\sin
2\phi(x))$ at the $x$ axis (see Fig.3b), where $c$ is a constant.
It is obvious that $\nabla \times \vec{E}_T =0$, i.e. $\oint_C
\vec{E}_T \bullet d\vec{l} =0$. However, in general $\oint_C
\vec{E}_{\omega}\bullet d\vec{l}$ and $\oint_C \vec{E}_s\bullet
d\vec{l}$ are separately non-zero.

Finally, we estimate the magnitude of $\vec{E}_T$. We use
parameters consistent with realistic experimental samples. Take
the Rashba parameter $\alpha = 3 \times 10^{-11}$eVm
(corresponding to $k_R$=1/100nm for $m=0.036m_e$), $\rho_s=10^6$/m
(i.e. one moving electron per 1000 nm in length), and
$k=k_F=10^8$/m. The electric potential difference between the two
points A and B (see Fig.3b) is about $0.01\mu V$, where the
positions of A and B are $\frac{1}{2k_R}(\frac{\pi}{2},0,0.01)$
and $\frac{1}{2k_R}(\frac{\pi}{2},0,-0.01)$. This value of the
potential is measurable with today's technology.\cite{addref2}
Furthermore, with the above parameters the electric field
$\vec{E}_T$ at A or B is about $5$V/m which is rather large.

\section{Conclusion}
In summary, we find that in order to completely describe the spin
flow (including both classic and quantum flows), apart from
the conventional spin current (or linear spin current), one has to
introduce another quantity, the angular spin current. The angular
spin current describes the rotational motion of the spin, and it
plays a parallel role in comparison with the conventional linear
spin current for the spin translational motion. Moreover we point
out that the angular spin current can also induce an electric
field and its $\vec{E}_{\omega}$ field scales as $1/r^2$ at large
$r$. In addition, a conserved quantity, the current
$\vec{j}_{\nabla\bullet\vec{s}}$ of the spin divergence, is
discovered, and the total electric field only depends on
$\vec{j}_{\nabla\bullet\vec{s}}$.

\section*{Acknowledgments}
We gratefully acknowledge financial support from the Chinese
Academy of Sciences and NSFC under Grant No. 90303016 and No.
10474125. XCX is supported by US-DOE under Grant No.
DE-FG02-04ER46124, NSF CCF-0524673 and NSF-MRSEC under DMR-0080054.

\newpage
\begin{figure}

\caption{(Color online) (a) and (b) are the schematic diagram for
the translational motion and the rotational motion of the classic
vector $\vec{m}$, respectively. (c) Schematic diagram for a
classic vector flow. }\label{fig:1}

\caption{(Color online) (a) The linear spin current element
$j_{s,xy}$. (b) and (c) The angular spin current element
$j_{\omega,x}$. (d) The spin current in a quasi 1D quantum wire.
(e) The currents of two magnetic charges that are equivalent to a
angular MM current. }\label{fig:2}

\caption{(Color online) (a) Schematic diagram for the spin moving
along the $x$ axis, with the spin precession (rotational motion)
in the $x$-$y$ plane while $0<x<L$. (b) A 1D wire of electric
dipole moment $\vec{p}_e$. This configuration will generate an
electric field equivalent to the field from the spin currents in
(a).}\label{fig:3}

\end{figure}


\begin{references}
\bibitem[*] Electronic address: sunqf@aphy.iphy.ac.cn

\bibitem{ref1}
S. A. Wolf, D. D. Awschalom, R. A. Buhrman, J. M. Daughton, S. V.
Molnar, M. L. Roukes, A. Y. Chtchelkanova, and D. M. Treger,
Science {\bf 294}, 1488 (2001); G.A. Prinz, Science {\bf 282},
1660 (1998).

\bibitem{ref2}
I. Zutic, J. Fabian, and S. Das Sarma, Rev. Mod. Phys. {\bf 76}, 323
(2004).

\bibitem{ref3}
S. Murakami, N. Nagaosa, S.-C. Zhang, Science {\bf 301}, 1348
(2003); Phys. Rev. B {\bf 69}, 235206 (2004); Z.F. Jiang, R.D. Li,
S.-C. Zhang, and W.M. Liu, Phys.Rev. B {\bf 72}, 045201 (2005).

\bibitem{ref4}
J. Sinova, D. Culcer, Q. Niu, N.A. Sinitsyn, T. Jungwirth, and
A.H. MacDonald, Phys. Rev. Lett. {\bf 92}, 126603 (2004); B.K.
Nikolic, S. Souma, L.P. Zarbo, and J. Sinova, cond-mat/0412595
(2004).

\bibitem{ref5}
S.-Q. Shen, M. Ma, X.C. Xie, and F.C. Zhang, Phys. Rev. Lett. {\bf
92}, 256603 (2004); cond-mat/0410169 (2004); S. Murakami, N.
Nagaosa, and S.-C. Zhang, Phys. Rev. Lett. {\bf 93}, 156804
(2004); S.-P. Kou, X.-L. Qi, and Z.-Y. Weng, cond-mat/0412146
(2004).

\bibitem{yao}
G.Y. Guo, Y. Yao, and Q. Niu, Phys. Rev. Lett. {\bf 94} 226601
(2005); Y. Yao and Z. Fang, cond-mat/0502351 (2005); X. Dai, Z.
Fang, Y.-G. Yao, and F.-C. Zhang, cond-mat/0507603 (2005).

\bibitem{ref6}
S. Datta and B. Das, Appl. Phys. Lett. {\bf 56}, 665 (1990).

\bibitem{ref7}
T. Matsuyama, C.-M. Hu, D. Grundler, G. Meier, and U. Merkt, Phys.
Rev. B {\bf 65}, 155322 (2002); F. Mireles and G. Kirczenow, {\sl
ibid}, {\bf 64}, 024426 (2001).

\bibitem{nn1}
P. Zhang, J.R. Shi, D. Xiao, and Q. Niu, Cond-mat/0503505 (2005);
J. Wang, B.G. Wang, W. Ren, and H. Guo, Cond-mat/0507159 (2005).


\bibitem{niu}
D. Culcer, J. Sinova, N.A. Sinitsyn, T. Jungwirth, A.H. MacDonald,
and Q. Niu, Phys. Rev. Lett. {\bf 93}, 046602 (2004).

\bibitem{zhang}
S. Zhang and Z. Yang, Phys. Rev. Lett. {\bf 94} 066602 (2005).

\bibitem{addref1}
A.A. Burkov, A.S. Nunez, and A.H. MacDonald, Phys. Rev. B {\bf
70}, 155308 (2004).

\bibitem{nnote1}
If to consider that there are many particles in the volume
element $\Delta V$ (i.e. the volume element $\Delta V$ is very
small macroscopically but very large microcosmically),
the vector direction $\vec{m}_i$, the velocity $\vec{v}_i$, and
the angular velocity $\vec{\omega}_i$ for each particle may be
different, however, the vector continuity equation (3) is still
valid:
 $
 \frac{d}{dt} \vec{M} =   - \nabla \bullet {\bf j}_s
  + \vec{j}_{\omega},
 $
with  ${\bf j}_s({\bf r},t) \equiv <\vec{v}\vec{M}> =
\lim\limits_{\Delta V \rightarrow 0} \frac{\sum_{i}
\vec{v}_i\vec{m}_i}{\Delta V}$ and $\vec{j}_{\omega} ({\bf r}, t)
\equiv <\vec{\omega}\times\vec{M}>= \lim\limits_{\Delta V
\rightarrow 0} \frac{\sum_{i} \vec{\omega}_i \times
\vec{m}_i}{\Delta V}$. Here ${\bf j}_s({\bf r}, t)$ and
$\vec{j}_{\omega}({\bf r},t)$ still describe the translational
and the rotational motions of the classical vector.

\bibitem{nnote2}
The scalar (e.g. charge $e$) continuity equation $\frac{d}{dt}
\rho^e +\nabla \bullet \vec{j}_e =0$ is from the kinematics and
the invariance of the charge $e$. It is independent of the
external force $F$ as well dynamic laws. In other words, even
if the acceleration $a \not= F/m$, the continuity equation still
survives. It is complete same with the vector continuity
equation (2) or (3). This equation is also from the kinematics and
the invariance of $|\vec{m}|$. In particular, it is independent
of the external force and the torque acting on the vector, as
well as the dynamic laws. In other words, the vector continuity
equation does not depend on the changes in the velocity and the angular
velocity under the actions of the forces and the torques.


\bibitem{note1}
Note that those results, Eqs.(10, 12, and 13), are valid in
general. They are independent of the special choice of Hamiltonian
(6). For example, in the case with a vector
potential $\vec{A}$, the general spin-orbit coupling $\alpha
\hat{\vec{\sigma}}\bullet [\vec{p}\times\nabla V({\bf r})]$, and
so on, the results still hold.

\bibitem{ref8}
E.I. Rashba, Fiz. Tverd. Tela (Leningrad) {\bf 2}, 1224 (1960)
[Solid State Ionics {\bf 2}, 1109 (1960)].

\bibitem{ref9}
Y.A. Bychkov and E.I. Rashba, J. Phys. C {\bf 17}, 6039 (1984).

\bibitem{add1note1}
$\vec{j}_{\omega}$ is called the spin torque in the work\cite{niu}
in which the semi-classical approach to the spin continuity
equation is discussed. Here we consider that $\vec{j}_{\omega}$
describes the rotational motion of a spin, thus named the angular
spin current.

\bibitem{note2}
From ${\bf j}_s({\bf r}, t)$, the total linear spin current
along $i$ direction ($i=x,y,z$) is: $\vec{I}_{si}(i,t)=\iint dS
\hat{i}\bullet {\bf j}_s({\bf r}, t)$. To assume
$\vec{I}_{si}(i,t)$ independent on $t$ and $i$ (e.g. in the case
of the steady state and without spin flip), one has:
$\vec{I}_{si}=\frac{1}{L}\iiint dV \hat{i}\bullet {\bf j}_s({\bf
r},t) =\frac{1}{L}\iiint dV Re \Psi^{\dagger} \hat{v}_i
\hat{\vec{s}} \Psi = \frac{1}{L}\iiint dV \Psi^{\dagger}
\frac{1}{2}(\hat{v}_i \hat{\vec{s}} + \hat{\vec{s}} \hat{v}_i)
\Psi =<\frac{1}{2}(\hat{v}_i \hat{\vec{s}} + \hat{\vec{s}}
\hat{v}_i)>$, where $L$ is sample length in $i$-direction. This
definition is the same as in recent publications\cite{ref4}.

\bibitem{nnote3}
Here we consider another physical effect, the heat produced by the spin
currents. Assume a uniform isotropic conductor having a linear
spin current ${\bf j}_s$ and a charge current $\vec{j}_e$, and
considering the simple case that there exists no spin
flip process (i.e. $\vec{s}$ is conserved) so that
$\vec{j}_{\omega}=0$. Then the produced heat $Q$ in unit volume
and in unit time is $Q= \sum\limits_{i=x,y,z}
\frac{2\rho}{4}\left[ (|\vec{j}_{si}|+j_{ei})^2 + (|\vec{j}_{si}|
- j_{ei})^2 \right] =\rho \left( \sum\limits_{ij} j^2_{s,ij}
+\sum\limits_i (j_{ei})^2 \right) \equiv \rho ({\bf j}_s^2
+\vec{j}_e^2)$, where $\rho$ is the resistivity. So the produced heat
by the spin current can be expressed by ${\bf j}_s$ (for the
case of $\vec{j}_{\omega}=0$). Note the produced heat $Q$ depends
on ${\bf j}_s^2$, whereas the induced electric field
depends on $\nabla'\bullet {\bf j}_s -\vec{j}_{\omega}$.


\bibitem{ref10}
J.E. Hirsch, Phys. Rev. B {\bf 42}, 4774 (1990); {\bf 60}, 14787
(1999); F. Meier and D. Loss, Phys. Rev. Lett. {\bf 90}, 167204
(2003).

\bibitem{addref2}
F. Schutz, M. Kollar, and P. Kopietz, Phys. Rev. Lett. {\bf 91},
017205 (2003); Phys. Rev. B {\bf 69}, 035313 (2004).

\bibitem{ref11}
Q.-f. Sun, H. Guo, and J. Wang, Phys. Rev. B {\bf 69}, 054409
(2004).

\bibitem{griffiths} For example, see {\it Introduction to
Electrodynamics} (Prentice-Hall, Englewood Cliffs, NJ, 1989).

\bibitem{nnote4}
Notice that here $\vec{j}_{\nabla\bullet\vec{s}} \equiv \nabla'
\bullet {\bf j} -\vec{j}_{\omega}$, it is not $\nabla \bullet {\bf
j} -\vec{j}_{\omega}$. In the steady state, $\nabla \bullet {\bf
j} -\vec{j}_{\omega} =0 $, however $\vec{j}_{\nabla\bullet\vec{s}}
= \nabla' \bullet {\bf j} -\vec{j}_{\omega}$ is usually non-zero.

\end{references}
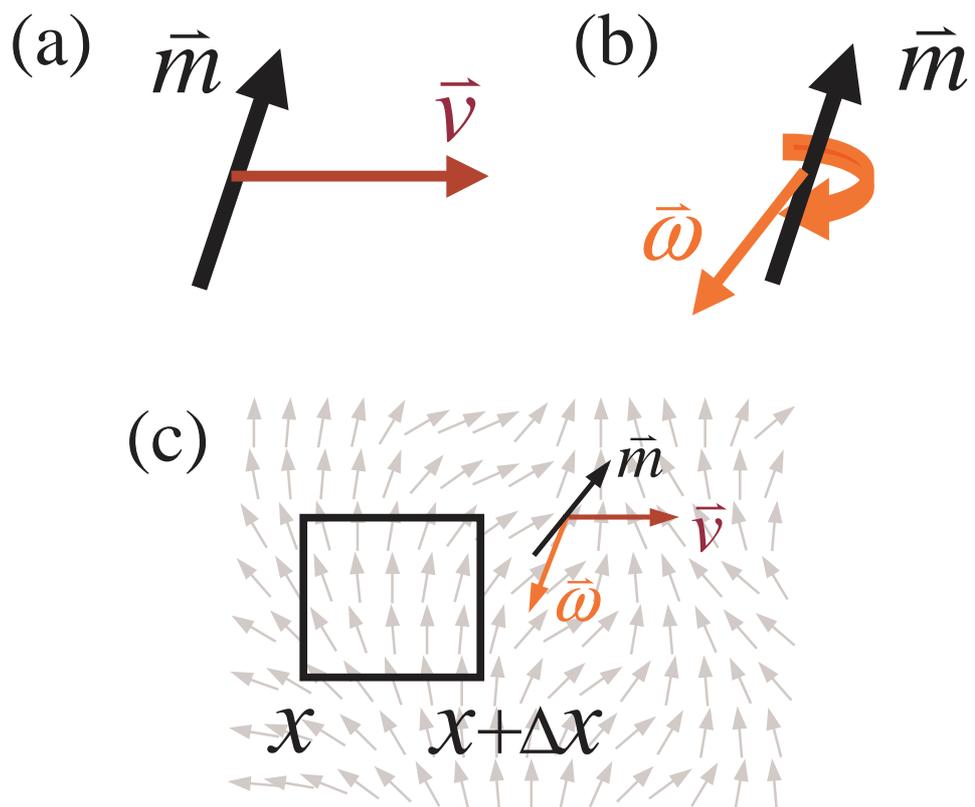
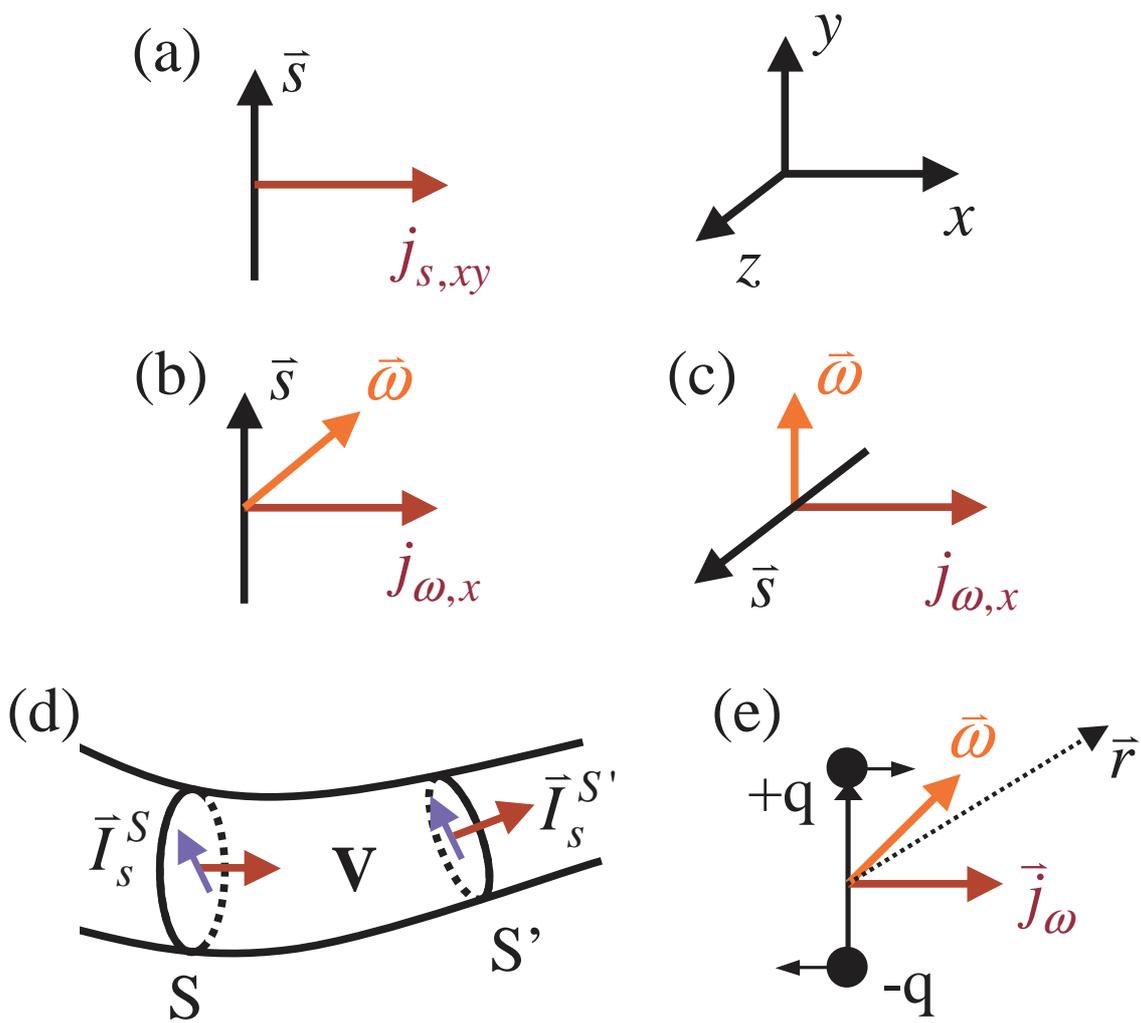
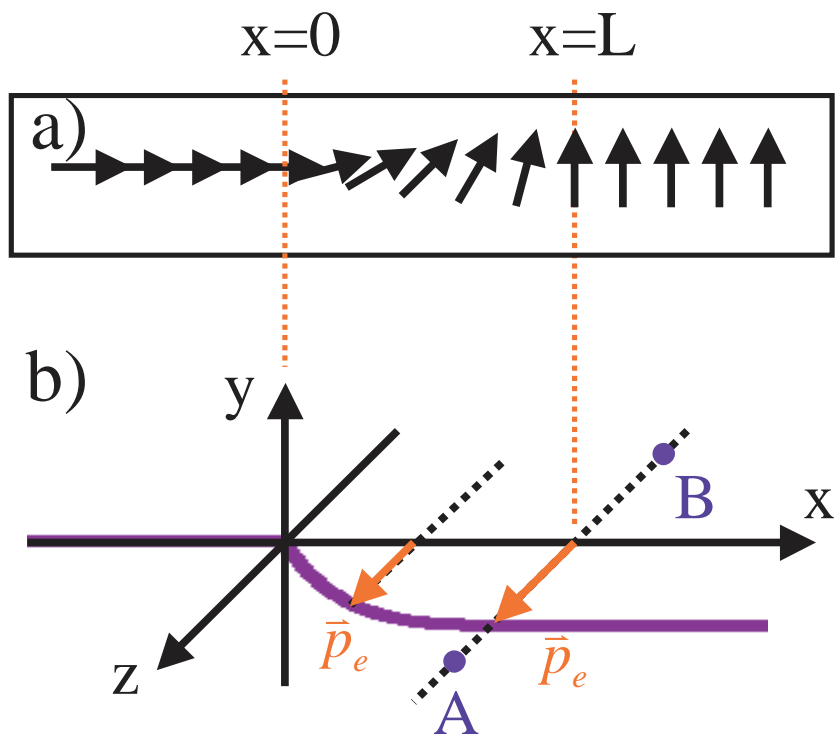
\end{document}